\documentclass[letter]{aa} 
\usepackage{graphicx}
\usepackage{natbib}
\usepackage{color}
\usepackage{upgreek}
\usepackage{soul}

\bibpunct{(}{)}{;}{a}{}{,}

\newcommand{\bz}{\langle B_{\rm z}\rangle}

\newcommand{\kms}{km\,s$^{-1}$}

\newcommand{\vsini}{v \sin i}

\usepackage{txfonts}
\usepackage{mathrsfs}
\usepackage[colorlinks=true, linkcolor=blue, citecolor=blue, filecolor=blue, urlcolor=blue]{hyperref}
\begin{document}

\titlerunning{Strong field of HD\,179218}
\authorrunning{S.~P.~J\"arvinen et al.}

\title{Detection of a large-scale organized 2\,kG order magnetic field in the 
Herbig\,Ae star HD\,179218}

\author{
	S.~P.~J\"arvinen\inst{1}
	\and
	S.~Hubrig\inst{1}
	\and
	M.~Sch\"oller\inst{2}
	\and
	I.~Ilyin\inst{1}
	\and
	H.~N.~Adigozalzade\inst{3}
	\and
	N.~Z.~Ismailov\inst{3}
	\and
	U.~Z.~Bashirova\inst{3}
	\and
	S.~A.~Alishov\inst{3}
                              }
\institute{
Leibniz-Institut f\"ur Astrophysik Potsdam (AIP), An der Sternwarte~16, 14482~Potsdam, Germany\\
 \email{sjarvinen@aip.de}
	\and
European Southern Observatory, Karl-Schwarzschild-Str.~2, 85748 Garching, Germany            
	\and
N.~Tusi Shamakhy Astrophysical Observatory, settl.~Y.~Mammadaliyev, 5626 Shamakhy, Azerbaijan
 }

\date{Received 30 April 2026 / Accepted 18 May 2026}

 
  \abstract
  {
While about two dozen Herbig Ae/Be stars have been reported to be magnetic,
only two of them, HD\,101412 and HD\,190073, have had their magnetic field 
geometries studied in the past. The knowledge of the magnetic field 
structure is important to understanding how magnetospheric accretion works 
in these stars.
  }
   {
We aim to study in detail the spectral and magnetic variability of 
HD\,179218, which is necessary to put constraints on its magnetic field 
geometry.
}
{
We measured the mean longitudinal magnetic field, $\bz$, from newly acquired and 
archival high-resolution spectropolarimetric observations of HD\,179218 using 
the least-squares deconvolution technique. Additionally, we studied the 
spectral variability of the hydrogen lines using dynamical spectra.
}
{
Based on our analysis of the Stokes~$V$ spectra of HD\,179218, we report for 
the first time the definite detection of a magnetic field. Using a slightly 
refined rotation period of $P_{\rm rot} = 1.34102$\,d, we constrained its 
geometry as follows: an estimated magnetic obliquity angle of 
$\beta=79.9\pm 0.7^{\circ}$ and a dipole strength of 
$B_{\rm d} = 2142\pm52$\,G. The $\bz$ variation is best fitted by the 
superposition of a sine wave and of its first harmonic, but more 
spectropolarimetric observations are necessary to test the impact of the 
limited measurement precision and the uneven coverage of the rotation cycle. 
The strongest emission in the H$\alpha$ and H$\beta$ line profiles in the 
medium-resolution spectra acquired in 2025 was detected close to the phases 
of the best visibility of the magnetic poles. HD\,179218 is the second 
Herbig Ae/Be star after HD\,190073 for which a first snapshot of a 
magnetosphere is presented.
}
   {}

   \maketitle
%

\section{Introduction}\label{sect:intro}

Observations of the disk properties of intermediate mass Herbig Ae/Be stars 
suggest a close parallel to T\,Tauri stars. Herbig Ae/Be stars and classical 
T\,Tauri stars are surrounded by active accretion disks, and (probably) most 
of the excess emission seen at various wavelength regions can be attributed 
to the interaction of the disk with a magnetically active star 
\citep[e.g.][]{Muzerolle2004}.
This interaction is generally referred to as magnetospheric accretion (MA). 
However, T\,Tauri stars generally have kilogauss-order complex magnetic field 
structures, while Herbig Ae/Be stars have been reported to possess much simpler
and far weaker fields, frequently below 100\,G
\citep[e.g.][]{Hubrig2009,Hubrig2011,Hubrig2015}.
While about two dozen Herbig Ae/Be stars are reported to be magnetic, only 
two Herbig Ae/Be stars, \object{HD\,101412} and \object{HD\,190073}, have 
had their magnetic field geometries constrained. These two stars have rather 
long rotation periods ($P_{\mathrm{rot}}$), with $42.076\pm0.017$\,d for 
HD\,101412 and $51.70\pm0.06$\,d for HD\,190073. It has been shown that 
these Herbig Ae/Be stars exhibit single-wave variations of the longitudinal 
magnetic field during the stellar rotation cycle. This behaviour is usually 
considered as evidence of a dominant dipolar contribution to the magnetic 
field topology. Among the studied stars, the Herbig Ae star HD\,101412 
possesses the strongest magnetic field, with the measured mean magnetic 
field modulus varying from 2.5 to 3.5\,kG 
\citep[e.g.][]{Hubrig2010}.

To understand how MA works in Herbig Ae/Be stars, knowledge of the magnetic 
field structure is indispensable. However, most Herbig Ae/Be stars are faint 
and require a considerable amount of telescope time. Therefore, the 
multi-epoch spectrophotometric observations necessary to estimate their 
rotation period and the magnetic field strength distribution over the 
rotation period are rare. The most recent study of the Herbig Ae star 
HD\,190073 with a nearly face-on disk orientation carried out by 
\citet{Silva2025} 
demonstrated that even stars with rather weak longitudinal magnetic fields 
of the order of a few tens of gauss possess sizeable magnetospheres. 
The authors reported that for HD\,190073, with a maximum longitudinal magnetic 
field of only 34\,G, 2D magnetohydrodynamical simulations show the presence 
of a compact magnetosphere with a radius of about $3\,R_*$ and a wind flow 
extending over tens of stellar radii ($R_*$).
\citet{Silva2025} 
used archival observations of HD\,190073 acquired by different 
spectropolarimeters from 2012 to 2019 and were able not only to determine the
rotation and magnetic period but also, for the first time, investigate in 
detail the remarkable variability of the emission Balmer line profiles of 
H$\alpha$, H$\beta$, and H$\gamma$ over the rotation period. The dynamical 
spectra constructed for the hydrogen line profiles revealed a ring-like 
magnetospheric structure appearing at the rotation phase of best visibility 
of the positive magnetic pole.

In this study we present our new spectropolarimetric observations of the 
Herbig Ae star \object{HD\,179218}. The star was extensively 
studied by 
\citet{Ismailov2024},
who reported the rotation period $P_{\mathrm{rot}}=1.341\pm0.002$\,d when 
using spectra obtained between 2015 with a resolution $R=14\,000$ and 2021 
($R=28\,000$) to study the variability of the H$\beta$ and the 
\ion{He}{i}~5876 lines. 
\citet{Cody2025}
found no periodicity in the Transiting Exoplanet Survey Satellite
\citep{Ricker-TESS}
observations. According to 
\citet{Fedele2008}, 
HD\,179218 has two rings of dust at 1 and 20 au and a compact gas emitting 
region between 1 and 6 au. 
\citet{Mandigutia2017} 
reported that MA is able to reproduce the bulk of the H$\alpha$ emission 
shown by HD\,179218, confirming previous estimates from MA shock modeling 
with a mass accretion rate of $10^{-8}M_{\sun}$\,yr$^{-1}$ and an 
inclination to the line of sight between $30^{\circ}$ and $50^{\circ}$. Most 
recent interferometric observations with GRAVITY indicate a disk inclination 
to the line of sight of $i=54^{+8}_{-10}$ 
\citep{Gravity2024}.

The first attempt to measure the magnetic field of HD\,179218 was carried 
out by
\citet{Hubrig2009}, 
based on low-resolution FOcal Reducer low dispersion Spectrograph
\citep[FORS1;][]{FORS1} 
observations on a single epoch. However, the measurement accuracy was too low
to be able to announce the detection of a magnetic field. A further 
attempt was made by
\citet{Alecian2013b}, 
this time using the high-resolution Echelle SpectroPolarimetric Device for the
Observation of Stars 
\citep[ESPaDOnS;][]{Espadons} 
and Narval
\citep{Narval}
spectropolarimeters on four observing epochs. Again, no definite detection 
was reported. These observations are publicly available from the 
Canada-France-Hawaii Telescope (CFHT) and PolarBase 
\citep{Petit2014}
data archives. 

To clarify the magnetic nature of HD\,179218, we obtained 
seven additional high-resolution spectropolarimetric observations in 2022 
and 2025 with the High Accuracy Radial velocity Planet Searcher polarimeter 
\citep[HARPS\-pol;][]{HARPS}
attached to the 3.6\,m European Southern Observatory (ESO) telescope at the La 
Silla Observatory in Chile. In Sect.~\ref{sect:obs} we describe the 
observational material used in this work, and we present the results of our 
magnetic field measurements using our own and archival data, the obtained 
constraints on the magnetic field geometry, and the appearance of dynamical 
spectra constructed for the H$\alpha$, H$\beta$, and \ion{He}{i}~5876 lines. 
In the Sect.~\ref{sect:disc} we discuss our new results in the light of the 
current knowledge of magnetism among Herbig Ae/Be stars and new possible 
directions in their studies.


\section{Observations and magnetic field measurements}
  \label{sect:obs}

\begin{table*}
\caption{
Logbook of the high-resolution spectropolarimetric observations and the 
results of the magnetic field measurements. 
}
\label{tab:obsall}
\centering
\begin{tabular}{ccccccc r@{$\pm$}l }
\hline\hline \noalign{\smallskip}
\multicolumn{1}{c}{Inst} &
\multicolumn{1}{c}{MJD} &
\multicolumn{1}{c}{$\varphi$} &
\multicolumn{1}{c}{$S/N$} &
\multicolumn{1}{c}{Line}&
\multicolumn{1}{c}{FAP}&
\multicolumn{1}{c}{Det.} &
\multicolumn{2}{c}{$\left< B \right>_{\rm z}$} \\
\multicolumn{1}{c}{} &
\multicolumn{1}{c}{} &
\multicolumn{1}{c}{} &
\multicolumn{1}{c}{} &
\multicolumn{1}{c}{mask} &
\multicolumn{1}{c}{} &
\multicolumn{1}{c}{flag} &
\multicolumn{2}{c}{[G]} \\
\noalign{\smallskip}\hline \noalign{\smallskip}
E & 53422.649074 & 0.153 & 177 & MgSiTiCrFe & $9\times10^{-4}$ & MD & $-$68  & 65  \\
E & 53608.363169 & 0.641 & 396 & CrFe       & $5\times10^{-7}$ & DD & 173    & 63  \\
N & 55107.912442 & 0.856 & 413 & MgSiTiCrFe & $1\times10^{-4}$ & MD & 447    & 66  \\
H & 59693.387609 & 0.250 & 261 & SiCrFe     & $2\times10^{-7}$ & DD & $-$117 & 21  \\
H & 59699.406061 & 0.738 & 220 & MgSiFe     & $9\times10^{-7}$ & DD & 343    & 55  \\
H & 60861.141508 & 0.045 & 279 & SiCaTiCrFe & $8\times10^{-8}$ & DD & $-$29  & 38  \\
H & 60861.302527 & 0.165 & 216 & TiCrFe     & $2\times10^{-5}$ & MD & $-$94  & 96  \\
H & 60862.290035 & 0.902 & 191 & SiTiCrFe   & $1\times10^{-7}$ & DD & 418    & 73  \\
H & 60863.272849 & 0.635 & 242 & TiCrFe     & $2\times10^{-4}$ & MD & 202    & 106 \\
H & 60865.257535 & 0.115 & 290 & CaSiTiCrFe & $3\times10^{-5}$ & MD & $-$69  & 38  \\
 \hline 
\end{tabular}
\tablefoot{
The first column shows the instrument used: `E' indicates ESPaDOnS, `N' Narval, 
and `H' HARPS\-pol. This is followed by the modified Julian date (MJD) 
values at the middle of the exposure, the corresponding phase (MJD for phase 
zero $T_{0}=59\,693.052435$, $P_{\mathrm{rot}}=1.34102$\,d), and the 
signal-to-noise ratio measured in the Stokes~$I$ spectra in the spectral 
region around 5000\,\AA. The line mask used, the FAP values, the detection 
flag (with `DD' meaning `definite detection' and `MD' standing for `marginal 
detection'), and the measured LSD mean longitudinal magnetic field strength 
are presented in Columns 5--8.
}
\end{table*}

The spectropolarimetric observations analysed in this paper were obtained on 
April~24 and 30, 2022, as well as on July~5 to~9, 2025, with HARPS\-pol. The 
target was also observed in spectropolarimetric mode with ESPaDOnS, installed 
at the CFHT, during February~21 and August~26, 2005, and with Narval installed 
at the Bernard Lyot Telescope at the Pic du Midi, France, during 
September~15, 2007, and October~3, 2009. However, the exposure time for the 
first Narval observation was only 600\,s, making these data unusable for our 
analysis. HARPS\-pol provides a resolving power of about 115\,000, whereas 
for ESPaDOnS and Narval the spectral resolution is about 65\,000. A summary 
of all observations is presented in Table~\ref{tab:obsall}.

HD\,179218 has also been intensively observed at medium resolution with the 
2\,m telescope of the Shamakhy Astrophysical Observatory. We used the two
most complete datasets to monitor the variability of the H$\alpha$, 
H$\beta$, and \ion{He}{i}~5876 lines with dynamical spectra. The first data 
set consists of 28 individual spectra obtained between May and August 2015 
with a resolution $R=14\,000$. The details of the observations and the setup 
are given in 
\citet{Ismailov2019}.
The second dataset used in our study was obtained between June and August 
2025, with a total of 45 spectra at $R=28\,000$. The details of these 
observations are presented in Table~\ref{tab:obsmed} in the Appendix.

\begin{figure}
    \centering
    \includegraphics[width=.75\columnwidth]{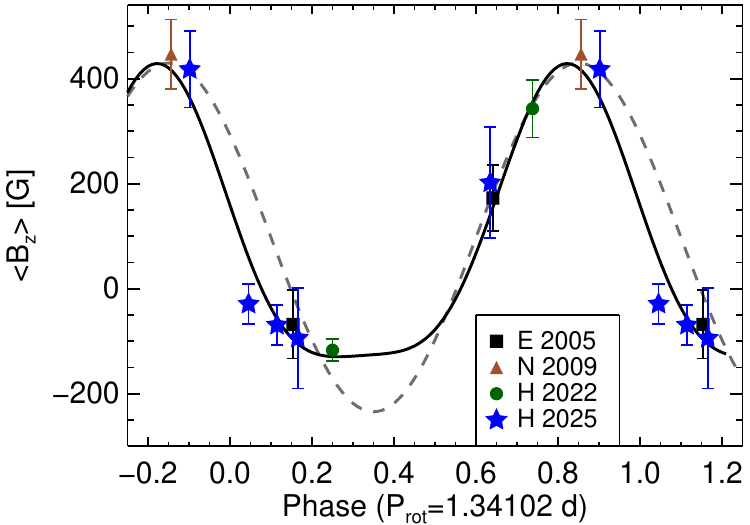}
    \caption{
Magnetic phase curve for HD\,179218. The measured 
$\left< B_{\rm z} \right>$ values presented in Table~\ref{tab:obsall} have been 
phased with the refined period $P_{\rm rot}=1.34102$\,d. Different symbols with 
different colours refer to observations acquired with ESPaDOnS in 2005,
Narval in 2009, and HARPS\-pol in 2022 and 2025. The dashed line is the 
sinusoidal fit to the data points, assuming a purely dipolar field 
configuration, whereas the solid line includes a contribution of the first 
harmonic.
   }
    \label{fig:Bzphase}
\end{figure}

\begin{figure*}
    \centering
    \includegraphics[width=0.28\textwidth]{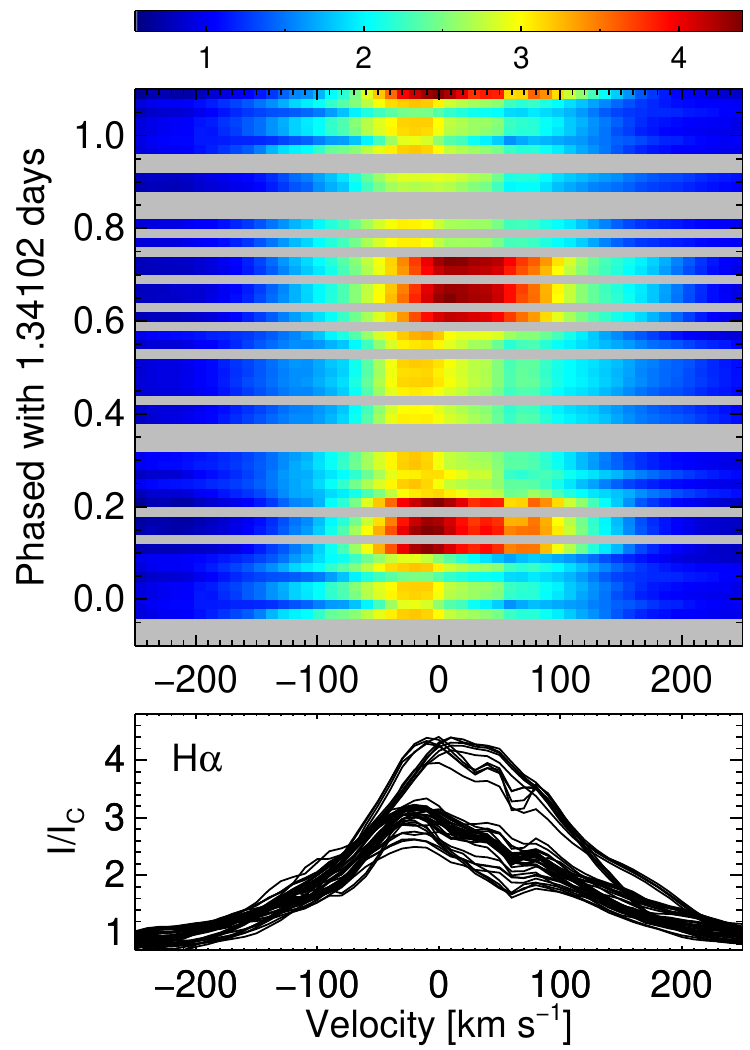}
    \includegraphics[width=0.28\textwidth]{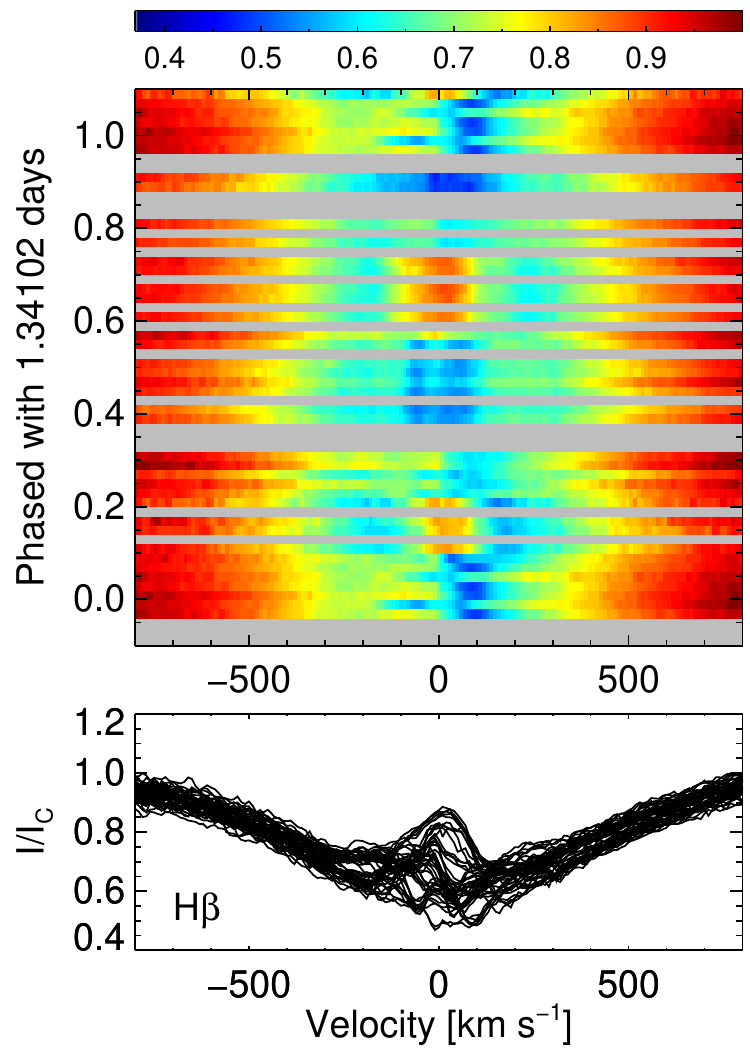}
   \includegraphics [width=0.28\textwidth]{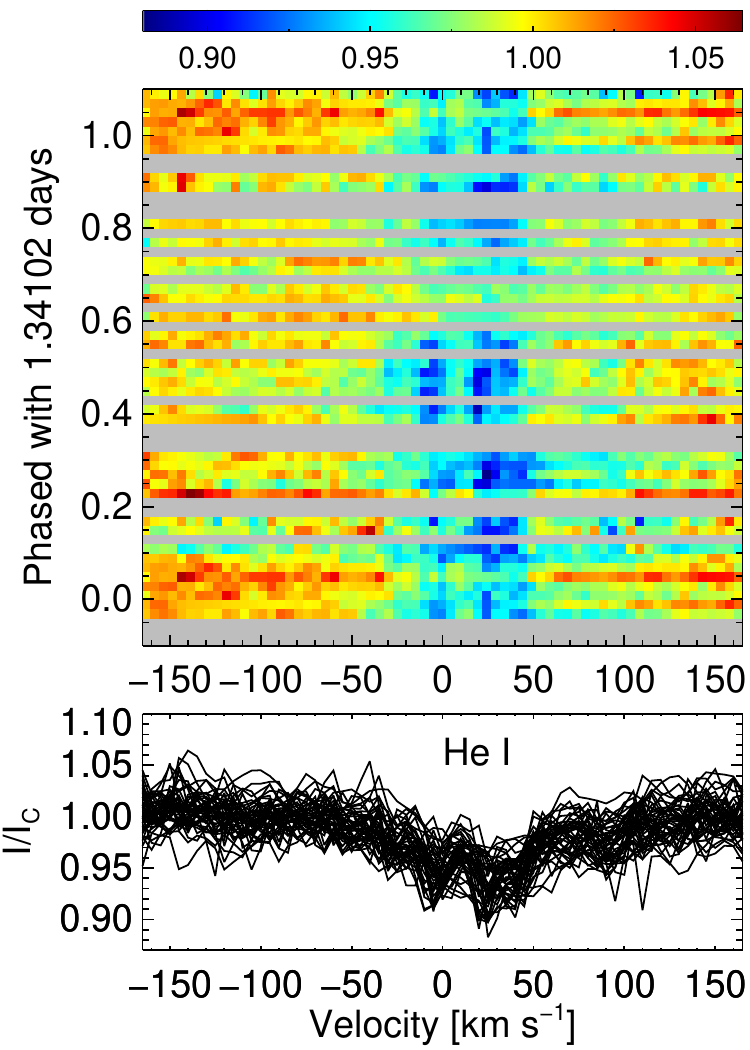}
    \caption{
Dynamical spectra constructed for the H$\alpha$ (left), H$\beta$ (middle), 
and \ion{He}{i}~5876 (right) line profiles. They are based on medium resolution 
spectra obtained in 2025 and are phased with $P_{\mathrm{rot}} = 1.34102$\,d.
The overplotted individual line profiles are shown in the bottom panels.
   }
    \label{fig:HaHb2025}
\end{figure*}

The mean longitudinal magnetic field $\bz$, accessible through 
high-resolution spectropolarimetric observations, is the line-of-sight 
component of the magnetic field weighted with the line intensity and 
averaged over the visible hemisphere. It is strongly dependent on the 
viewing angle between the field orientation and the observer and is 
modulated as the star rotates. As in our previous studies
\citep[e.g.][]{Silva2025,Hubrig2025},
we employed the least-squares deconvolution (LSD) technique following the 
description given by 
\citet{Donati1997LSD}.
This technique allows the accuracy of the mean longitudinal magnetic field 
determination to be increased. The parameters for the individual lines are 
based on information provided by the Vienna Atomic Line Database 
\citep[VALD3;][]{VALD}
using an effective temperature $T_{\mathrm{eff}} = 9500 \pm 250$\,K and 
a surface gravity $\log g = 3.95 \pm 0.10$
\citep{Wichittanakom2020} 
and a detection threshold of 0.1. To evaluate the detected features, we 
followed the false alarm probability (FAP) based on reduced $\chi~^2$ test 
statistics 
\citep{Donati1992FAP}:
The presence of the Zeeman feature is considered a definite detection 
(DD) if FAP $\leq 10^{-5}$, a marginal detection (MD) if 
$10^{-5} <$ FAP $\leq 10^{-3}$, and a non-detection (ND) if FAP 
$> 10^{-3}$.

The results of our magnetic field measurements and the corresponding FAP values 
are given in Table~\ref{tab:obsall}, while the obtained LSD profiles ordered 
by increasing rotation phases are shown in Appendix~\ref{App:IVN} in 
Fig.~\ref{fig:IVN}. Since the data used were obtained over 20 years and 
the rotation period of 1.341\,d is very short and not especially accurate, 
the measured $\bz$ strengths plotted against the phase 
displayed noticeable dispersion with phase shifts between similar field 
strength values. After the refinement of the period and taking into account the 
phase shift between the field strength measured on August~26, 2005, and the 
measurements on July~7, 2025, we obtained a period of 1.34102\,d. In 
Fig.~\ref{fig:Bzphase} we display the $\bz$ plotted against the rotation 
phase and fitted by a sine curve (the dashed line), under the assumption 
that the field geometry is a pure dipole. Reasonable evidence that the field 
geometry is predominantly dipolar follows from the appearance of the 
dynamical spectra constructed for the H$\alpha$ and H$\beta$ line profiles 
observed using the medium-resolution dataset from 2025. As presented in 
Fig.~\ref{fig:HaHb2025}, the hydrogen line profiles show a clear increase 
of emission twice over the rotation cycle. As recently discussed by 
\citet{Silva2025} 
and 
\citet{Hubrig2025}, 
such an increase is related to a stellar magnetosphere and generally occurs
when one of the magnetic poles comes into sight as the star rotates. Also, 
other spectral lines exhibit a distinct variability. In 
Fig.~\ref{fig:HaHb2025} we show that despite the \ion{He}{i}~5876 line 
profiles being rather weak and noisy, they generally follow the behaviour of 
the H$\beta$ profiles. At some phases both lines display a double-absorption 
structure in the line cores, best visible around the phase 0.5. It is not 
clear whether these absorptions are related to the surrounding hot inner disk 
region mentioned by 
\citet{Kokoulina2021}.
In addition, in Fig.~\ref{fig:elem} in the Appendix we show the 
combined variability of Ti, Cr, and Fe lines. In contrast to the observations 
acquired in 2025, the emission strengths in the H$\alpha$ and  H$\beta$ lines 
observed in 2015 and displayed in Fig.~\ref{fig:HaHb2015} are much lower 
and have no clear periodical changes in their line profiles. An annual 
change in the hydrogen line profiles was previously observed in HD\,190073 by
\citet{Silva2025}. 
The authors speculated that this was due to an interaction with a secondary 
body in the system causing an increase in the wind outflow and producing 
a partial screening of the magnetosphere.

Using the strongest measured $\bz{}$ of $447\pm 66$\,G and employing the 
relation $B_{\rm d} \ge 3 \left| \bz \right|$
\citep{Babcock1958},
we estimate the minimum dipole strength, $B_{\rm d}$, to be about 1.34\,kG.
Assuming a rotation period $P_{\rm rot}$ of 1.34102\,d and a stellar radius 
$R_\ast$ of $3.59 \pm 0.10\,R_\odot$ 
\citep{GuzmanDiaz2021},
we obtained an equatorial velocity of 
$v_{\rm eq} = 135.46 \pm 3.77$\,km\,s$^{-1}$. Using 
$\vsini = 68.8 \pm 2.9$\,\kms{} from the study of
\citet{Alecian2013b}, 
the angle between the rotation axis and the line of sight $i$ is 
$30.5 \pm 1.7^{\circ}$. The general description for the strength of the 
observed longitudinal magnetic field for a simple centred dipole was 
presented by
\citet{Preston1967}.
The relative amplitude of variation of the fitted longitudinal magnetic
field phase curve is usually characterised by the parameter $r$, representing
the ratio between $\bz_{\rm min}$ and $\bz_{\rm max}$. Here, assuming a
dipolar field, with
$\bz_{\rm max}=430\pm4$\,G and $\bz_{\rm min}=-234\pm4$\,G, we have
$r=-0.54 \pm 0.10$. Using
\begin{equation}
r  =  \frac{\bz_{\rm min}}{\bz_{\rm max}}
=  \frac{\cos \beta \cos i - \sin \beta \sin i}{\cos \beta \cos i + \sin \beta \sin i}
\label{eq:diagn.oblique_r},
\end{equation}
we calculated an obliquity angle of $\beta=79.9\pm 0.7^{\circ}$. With a
limb-darkening coefficient of 0.5 
\citep{Claret2019},
we obtained a polar magnetic field strength of $B_{\rm d}=2142\pm52$\,G. 

As demonstrated in Fig.~\ref{fig:Bzphase}, the $\bz$ phase curve shows a 
degree of anharmonicity that requires a contribution of a first harmonic. The 
solid line in this plot represents a fit of the observations by a 
superposition of a dipole and a quadrupole field at the centre of the star. 
Clearly, more measurements of $\bz$ are necessary to be able to decide 
whether it is possible that the anharmonic variation curve actually results 
from some combination of the limited precision of the measurements and 
their uneven coverage over the rotation cycle.


\section{Discussion}
\label{sect:disc}

HD\,179218 is the third Herbig Ae/Be star with a known rotation period, and 
it hosts the second strongest magnetic field (after that of HD\,101412), 
with a 2\,kG order dipolar component. A rotation period of 1.341\,d was 
estimated from the variable position of the red boundary of the absorption 
component of the inverse P\,Cyg-type profile of H$\beta$ by 
\citet{Ismailov2024}, 
who assumed that this variability is caused by MA. A similar procedure using 
structural kinematic features of the gas environment was previously 
successfully applied using near-infrared observations of the strongly 
magnetic Herbig Ae star HD\,101412 
\citep{Schoeller2016}. 
It would be worthwhile to apply such a procedure to other Herbig Ae/Be stars 
to estimate their rotation periods. However, since Herbig Ae/Be stars are 
complex systems with surrounding protoplanetary disks and diverse 
interactions between star, gas, and dust, extensive time series over 
hours, days, and months are necessary to disentangle the contributions from 
the different system components.

It is notable that based on the variability of the Balmer line profiles and 
the colour indices in the $U$ $BV$ $RI$ system monitored over several years,
\citet{Ismailov2025}  
also suggested the presence of longer periods from 37.5 to 737\,d, which 
may be explained by the presence of a distant companion and, possibly, one 
or more distant exoplanets around HD\,179218. The dynamical spectra of the
H$\alpha$ and H$\beta$ line profiles phased with the period of 37.5\,d are 
presented in Fig.~\ref{fig:HaHb375d} in the Appendix. Even if the phase 
coverage is poor, some hints of a structure with maximum emission in phases 
between 0.15 and 0.4 can be observed. Also the absence of periodicity over 
the 1.341\,d rotation period in the Balmer line profiles in the spectra 
acquired in 2015 suggests that some kind of dynamical process (probably 
due to the presence of a companion) takes place around the central star. 
According to 
\citet{Thomas2023}, 
who used adaptive optics infrared imaging surveys, one of the possible 
companions, HD\,179218\,B, is located at about 2.5\,\arcsec{}. That is too far 
to have an important dynamical impact, and no close companion was detected 
using interferometric observations. MATISSE L-band observations of 
HD\,179218 by 
\citet{Kokoulina2021} 
reported on the presence of two separate dust populations: a region filled 
with stochastically heated, very small (nano-)carbon grains within 10\,au and 
a colder outer disk for which the infrared emission is dominated by larger, 
probably micrometre-sized, grains.

The presented magnetic phase curve of HD\,179218 based on the longitudinal 
magnetic field measurements carried out on ten observing epochs indicates a 
more complex magnetic topology than a pure dipole and suggests a contribution 
of a quadrupolar field component. It is not clear whether this more complex 
field structure, if confirmed by future observations, is due to the faster 
rotation of this star in comparison to the rather slow rotation of the two 
previously studied Herbig Ae/Be stars (HD\,101412 and HD\,190073) with dipole 
fields. Obviously, additional spectropolarimetric observations of HD\,179218 
with a dense coverage of the rotation cycle will be valuable to ascertaining 
the structure of its magnetic field in more detail.


\begin{acknowledgements}

We thank the referee, P.\ Petit, for his comments.
This work is based on observations made with ESO telescopes at the La Silla 
Paranal Observatory under programme IDs 0109.C-0265(A) and 0115.D-2108(A) 
publicly available from the ESO Archive, on observations collected at the 
Canada-France-Hawaii Telescope (CFHT), which is operated by the National 
Research Council of Canada, the Institut National des Sciences de l'Univers 
of the Centre National de la Recherche Scientifique of France, and the 
University of Hawaii, and on observations collected at the Bernard Lyot 
Telescope at the Pic du Midi de Bigorre, France, which is managed by the 
Observatoire Midi Pyr\'en\'ees.

\end{acknowledgements}

%
   \bibliographystyle{aa} 
   \bibliography{hd179218} 
%

\begin{appendix}


\section{Medium resolution observations}\label{App:med}

The details of the medium resolution observations collected during 2025 are 
presented in Table~\ref{tab:obsmed}. The observations were obtained with the 
ShAO fiber echelle spectrograph (ShaFES). The spectrograph is described in
\citet{Mikailov2020}
and the reduction in
\citet{Ismailov2024}.

\begin{table}[!h]
\caption{
Logbook of the 2025 medium resolution observations. 
}
\label{tab:obsmed}
\centering
\begin{tabular}{ccccc }
\hline\hline \noalign{\smallskip}
\multicolumn{1}{c}{Date}&
\multicolumn{1}{c}{UT}&
\multicolumn{1}{c}{JD} &
\multicolumn{1}{c}{Exp}&
\multicolumn{1}{c}{$S/N$} \\
\multicolumn{1}{c}{} &
\multicolumn{1}{c}{} &
\multicolumn{1}{c}{2\,460\,000+} &
\multicolumn{1}{c}{[s]} &
\multicolumn{1}{c}{} \\
\noalign{\smallskip}\hline \noalign{\smallskip}
11.06.2025 & 18:03 & 838.2521 & 3600 & 76 \\
           & 19:19 & 838.3048 & 3600 & 74 \\
14.06.2025 & 20:35 & 841.3576 & 3600 & 68 \\
           & 21:25 & 842.3929 & 3600 & 66 \\
19.06.2025 & 19:32 & 846.3138 & 3600 & 76 \\
           & 20:32 & 846.3555 & 3600 & 76 \\
           & 21:33 & 846.3979 & 3600 & 75 \\
           & 22:33 & 846.4395 & 3600 & 79 \\
25.06.2025 & 19:29 & 852.3118 & 3600 & 65 \\
           & 20:30 & 852.3541 & 3600 & 63 \\
           & 21:31 & 852.3965 & 3600 & 62 \\
           & 22:32 & 852.4388 & 3600 & 63 \\
26.06.2025 & 19:23 & 853.2976 & 3600 & 62 \\
           & 20:23 & 853.3493 & 3600 & 69 \\
           & 21:24 & 853.3916 & 3600 & 66 \\
           & 22:25 & 853.4340 & 3600 & 63 \\
27.06.2025 & 19:27 & 854.3104 & 3600 & 69 \\
           & 20:23 & 854.3493 & 3600 & 60 \\
           & 21:23 & 854.3909 & 3600 & 69 \\
           & 22:28 & 854.4361 & 3600 & 65 \\
28.06.2025 & 19:07 & 855.2965 & 3600 & 65 \\
           & 20:07 & 855.3382 & 3600 & 63 \\
           & 21:08 & 855.3805 & 3600 & 68 \\
           & 22:09 & 855.4229 & 3600 & 66 \\
11.07.2025 & 18:30 & 868.2708 & 3600 & 84 \\
           & 19:33 & 868.3145 & 3000 & 81 \\
           & 20:23 & 868.3493 & 3000 & 83 \\
           & 21:13 & 868.3840 & 3000 & 82 \\
           & 22:04 & 868.4194 & 3000 & 80 \\
           & 22:54 & 868.4541 & 3000 & 81 \\
           & 23:44 & 868.4889 & 3000 & 83 \\
12.07.2025 & 21:55 & 869.4131 & 3000 & 79 \\
           & 22:45 & 869.4479 & 3000 & 78 \\
17.07.2025 & 19:28 & 874.3111 & 3000 & 76 \\
           & 20:18 & 874.3458 & 3600 & 76 \\
28.07.2025 & 19:51 & 885.3271 & 3600 & 61 \\
01.08.2025 & 20:49 & 889.3674 & 3600 & 60 \\
           & 21:50 & 889.4097 & 3600 & 59 \\
           & 22:50 & 889.4514 & 3600 & 61 \\
           & 23:50 & 889.4930 & 3600 & 62 \\
30.08.2025 & 16:40 & 918.1944 & 3600 & 65 \\
           & 17:41 & 918.2368 & 3600 & 64 \\
           & 18:41 & 918.2784 & 3600 & 60 \\
           & 19:41 & 918.3201 & 3600 & 66 \\
           & 20:42 & 918.3625 & 3600 & 63 \\
 \hline 
\end{tabular}
\tablefoot{
The first two columns give the date and universal time of the observations 
followed by the corresponding Julian date. The fourth column shows the 
exposure time and the fifth column gives the signal-to-noise ratio of the 
spectra.
}
\end{table}

\begin{onecolumn}

\section{LSD profiles}\label{App:IVN}

In Fig.~\ref{fig:IVN} we show the individual LSD profiles sorted according to
the phase given in Table~\ref{tab:obsall}.

\begin{figure*}[!h]
    \centering
    \includegraphics[width=\textwidth]{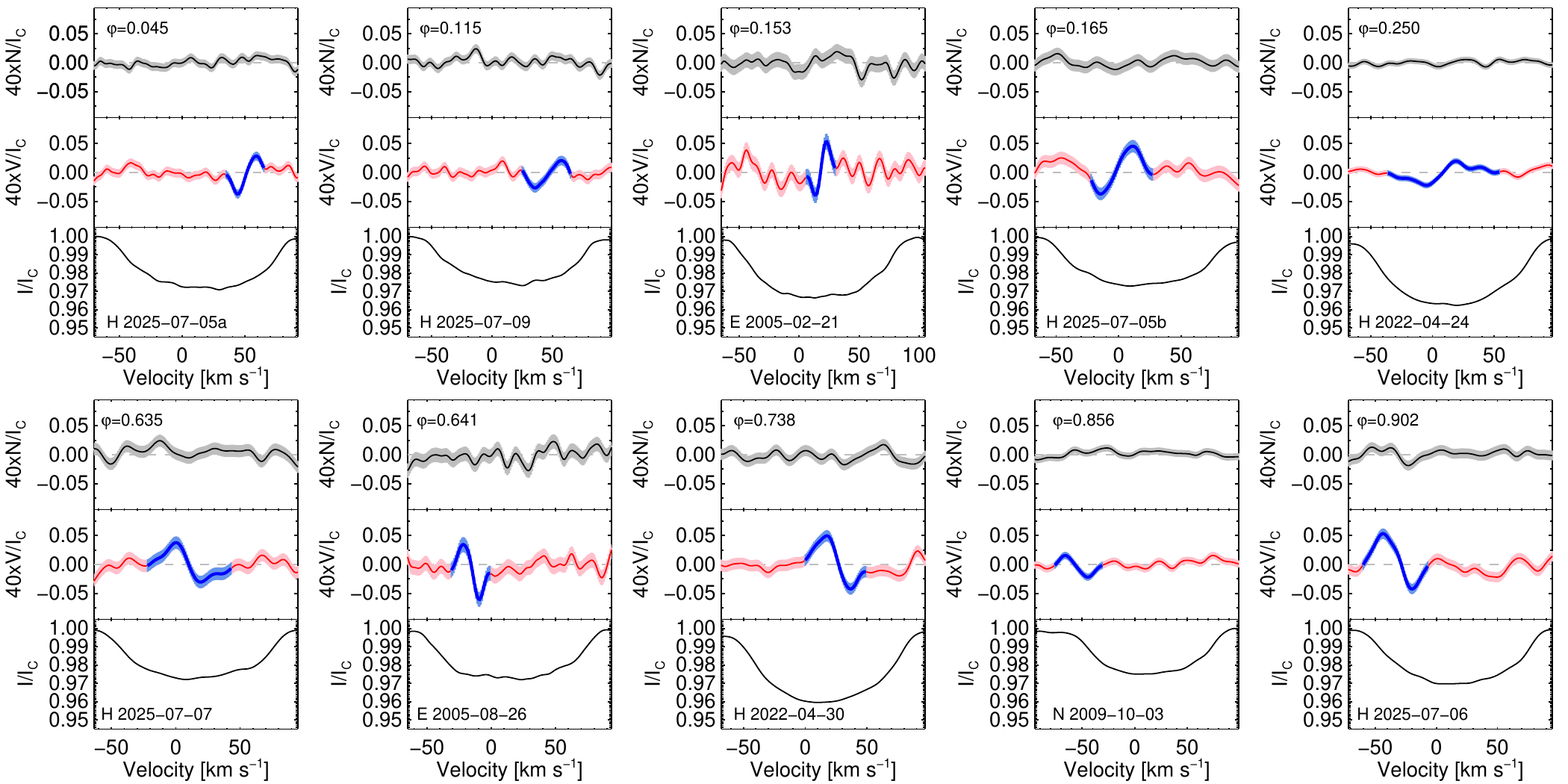}
    \caption{Least-squares deconvolution profiles. LSD Stokes~$I$ (bottom), Stokes~$V$ (middle), and 
diagnostic null (top) profiles for all epochs according to increasing 
phase. The Zeeman features are highlighted in blue.
   }
    \label{fig:IVN}
\end{figure*}


\section{LSD Stokes~$I$ profile variability observed for different elements}\label{App:var}

We show in Fig.~\ref{fig:elem} the variable LSD Stokes~$I$ profiles calculated 
using Ti, Cr, and Fe lines.

\begin{figure*}[!h]
    \centering
    \includegraphics[width=\textwidth]{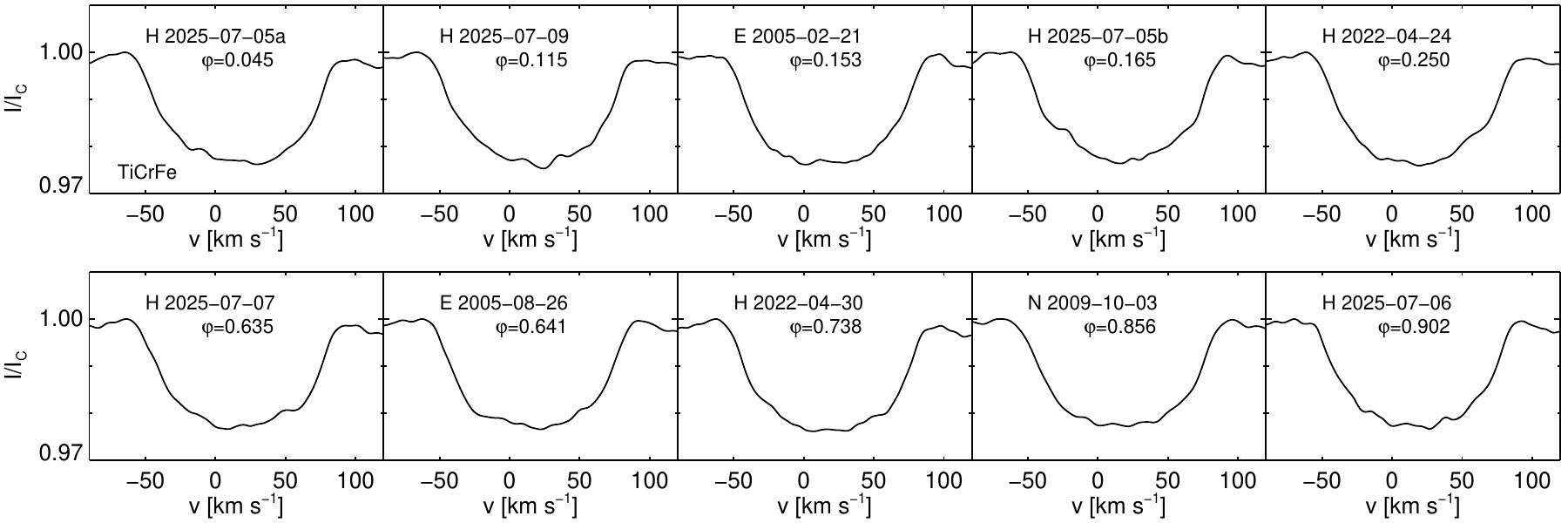}
    \caption{
 Least-squares deconvolution Stokes~$I$ profiles calculated using Ti, Cr, and Fe lines.
   }
    \label{fig:elem}
\end{figure*}

\end{onecolumn}


\begin{twocolumn}

\section{Line variability and additional periods}\label{App:per}

The H$\alpha$ profiles obtained in 2015 at $R=14\,000$ have been 
intensively analysed by
\citet{Ismailov2019}.
Neither H$\alpha$ nor H$\beta$ (dynamical spectrum and profiles visualised in
Fig.~\ref{fig:HaHb2015}) show significant structure. This might be caused by
some kind of dynamical process due to the presence of a companion.
\citet{Ismailov2025}
also suggested a longer period of $P_{\mathrm{rot}}=37.5$\,d based on the
analysis of H$\beta$ line variability. The dynamical spectrum using this 
period is presented in Fig.~\ref{fig:HaHb375d}. Albeit the phase coverage
is poor, some hints of a structure with a maximum emission in phases 
between 0.15 and 0.4 are detected.

\begin{figure}[!h]
    \centering
    \includegraphics[width=.45\columnwidth]{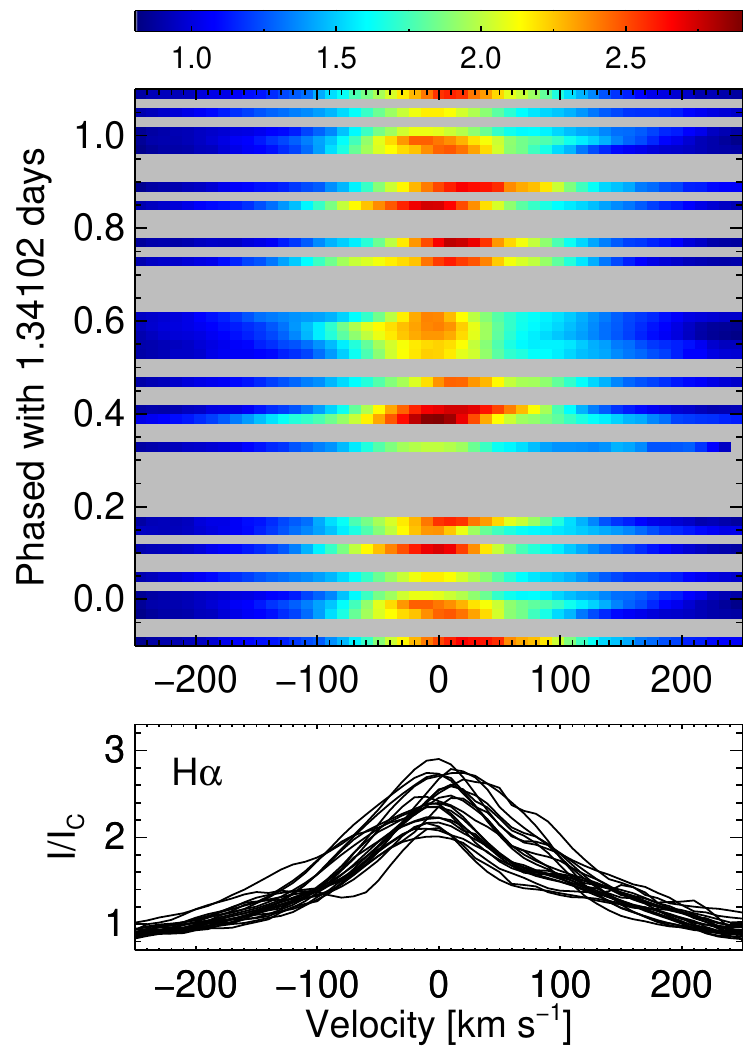}
    \includegraphics[width=.45\columnwidth]{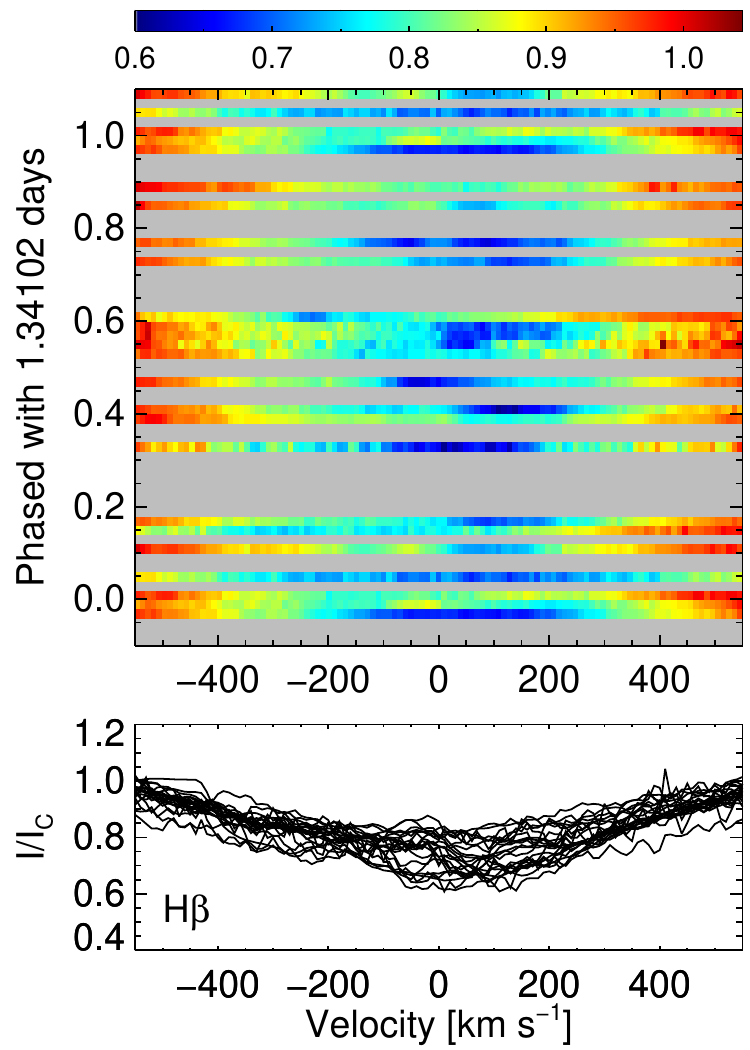}
    \caption{
Same as Fig.~\ref{fig:HaHb2025} but showing only H$\alpha$ and H$\beta$ based on the 2015
lower resolution spectra.
   }
    \label{fig:HaHb2015}
\end{figure}

\begin{figure}[!h]
    \centering
    \includegraphics[width=.45\columnwidth]{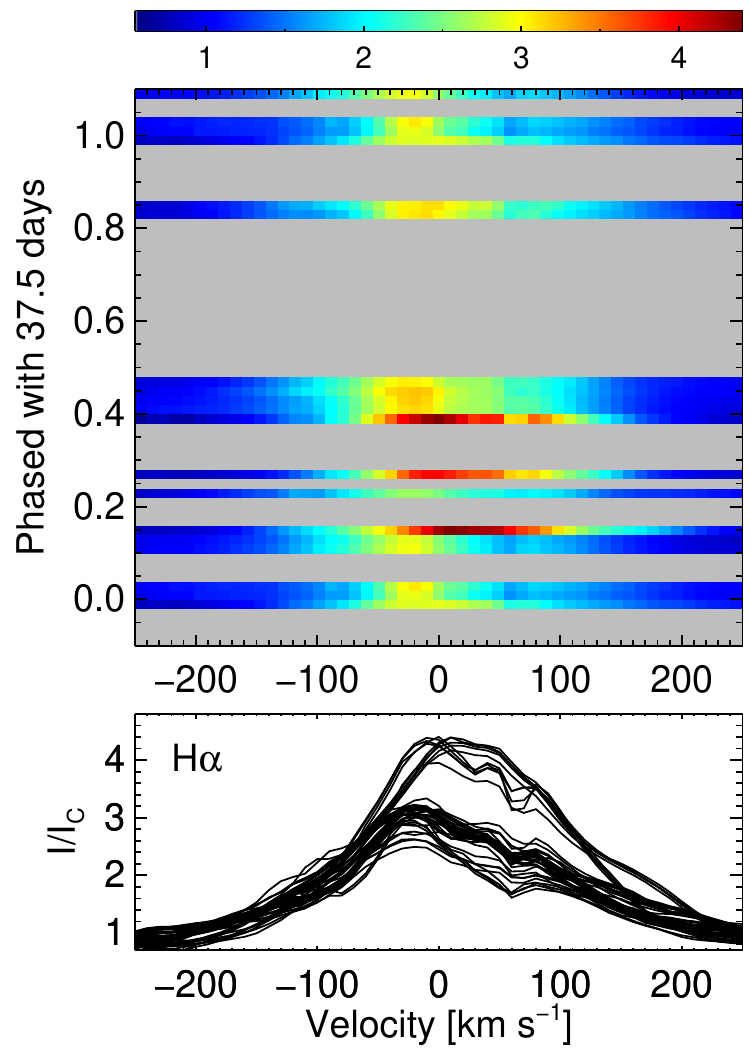}
    \includegraphics[width=.45\columnwidth]{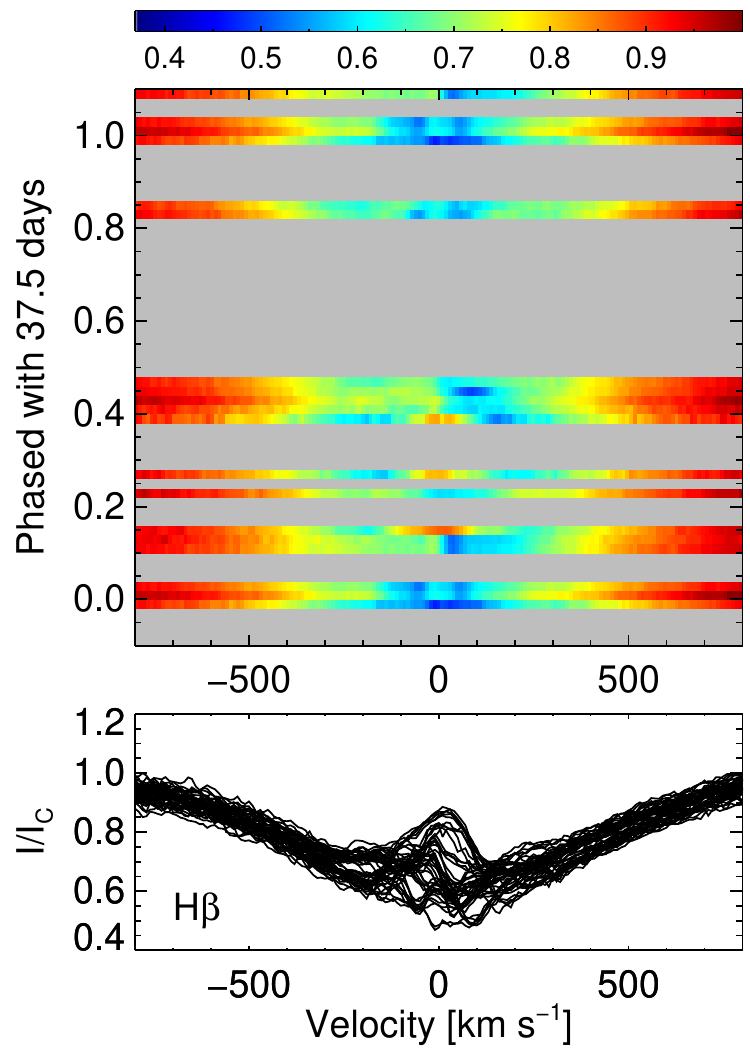}
    \caption{
Same as Fig.~\ref{fig:HaHb2025} but phased with $P_{\mathrm{rot}}=37.5$\,d, as
suggested by
\citet{Ismailov2025} and
based on H$\beta$ variability.
   }
    \label{fig:HaHb375d}
\end{figure}

\end{twocolumn}


\end{appendix}

\end{document}